# Datacenter Changes vs. Employment Rates for Datacenter Managers

In the Cloud Computing Era

Dr. Timur Mirzoev
Georgia Southern University
Statesboro, GA USA

David Hillhouse
Georgia Southern University
Statesboro, GA USA

Bruce Benson
Armstrong Atlantic University
Savannah, GA USA

Mickey Lewis
Southern Polytechnic State University
Marietta, GA USA

*Abstract*— Due to the evolving Cloud Computing paradigm, there is a prevailing concern that in the near future data center managers may be in short supply. Cloud computing, as a whole, is becoming more prevalent into today's computing world. In fact, cloud computing has become so popular that some are now referring to data centers as "cloud centers". How does this interest in cloud computing translate into employment rates for data center (or cloud center) managers? The popularity of the public and private cloud models are the prevailing force behind answering this question. Therefore, the skill set of the datacenter manager has evolved to harness the on demand self-services, broad network access, resource pooling, rapid elasticity, measured service, and multi tenacity characteristics of cloud computing [1]. Using diverse sources ranging from the Bureau of Labor and Statistics to trade articles, this manuscript takes an in-depth look at these employment rates related to the cloud and the determining factors behind them. Based on the information available, datacenter manager employment rates in the cloud computing era will continue to increase well into 2016.

## I. INTRODUCTION

How will the cloud affect employment rates of datacenter managers? Each year the global demand for data management is growing. For instance, the International Data Corporation (IDC) is predicting digital content to grow from just under a zettabyte to over 8 zettabytes by the end of 2015 [2]. This trend is creating interest in all cloud deployment models (private, public, community, and hybrid), and in Infrastructure as a Service (IaaS). Enterprises are craving reduced service level agreement (SLA's), reduced costs, and a more effective means to manage their digital content. The use of IaaS and all cloud deployment models have some researchers scared that it will reduce management jobs, since datacenters will be consolidated, make use of virtual machines, and automate resources.

The explosion of Big Data is fueling the need and sparking interest in cloud deployment models. Therefore, more and more datacenter managers will be needed with the experience to ascertain data and make effective decisions [3]. In addition, the McKinsey Global Institute is predicting the possibility of a shortage of 1.5 million skilled and qualified datacenter managers by 2015 [3]. All in all, cloud computing sparked by digital content growth will place a premium demand for managers with the required skills to build, operate, and maintain datacenters.

In the fourth quarter of 2010, IDC estimated over 19% of the physical servers shipped were configured to allow multiple customers running different operating systems to process at the same time [4]. This virtualization would allow many customers to use the same hardware as other customers. It was hard for them to judge how many IT organizations were performing trials with public and private clouds during the same time period. The increased virtualization has led to automation and an overall increase in flexibility on the data center. While these models are just being put into production on the IT side of the datacenter, the facilities side of the data center typically lags behind. Datacenter managers will have to be able to balance availability and uptime of their center while embracing new technologies. With the arrival of cloud computing, datacenter managers feel job security pressure especially if the datacenter in not up and running constantly [4].

A good example of datacenter managers feeling job pressure can be found at searchcio.techtarget.com [5]. In an August 15, 2012 article by features writer Karen Goulart, Ms Goulart stated concerning Seven Corners Inc. CIO George Reed, "having the right enterprise cloud computing skills in his data center is a matter of life and death" [5]. Taken into account the seriousness of the flow of information in today's society, it is certainly easy to understand that if information stops flowing, bad things can happen. This makes the availability of information held in "clouds", private or public, invaluable.

The biggest change required for the transition from in-house IT infrastructure to cloud computing is the shift to managing service, quality, and availability through contracts and relationships, rather than having an internal staff to manage the IT processes. But this implies a bigger corporate challenge: because different skills are required, existing IT staff will likely need to be retrained or replaced. Organizations will have to address the old issue: an attempt to transition to the cloud with existing staff will likely meet internal resistance, but new staff would lack knowledge of company processes. Overall, while more datacenter management positions will be created due to cloud technology, deeming a rise in the employment rates, the skill requirements will also place a demand and possible shortage of qualified datacenter managers.

The skill sets of datacenter managers have significantly changed with the evolution of the public and private cloud computing models. Furthermore, automation, via virtualized servers, has further developed via the use of all cloud-computing models, especially the public and private models. These data support the notion that there will be a shortage of datacenter managers due to the use of cloud computing via the popularity of the public and private cloud models.

## II. CURRENT TRENDS

To understand how employment of data center managers might be affected by cloud computing, some things that need to be understood are what some forms of cloud computing actually are, what the immediate trends are and what the long range forecast is for cloud computing as a whole. As mentioned in the introduction, cloud computing exists in many forms. There are both public and private clouds as well as hybrids [6]. Clouds are also classified by the type of service(s) provided.

Software as a Service (SaaS) offers the use of different software packages to the customer. Some examples of the software offered would be CRM or ERP software packages. This type of service relieves the customer from having to buy and maintain the software used. Infrastructure as a Service (IaaS), is another form of cloud computing. In this form of cloud, customers are most often connected to virtual machines. This allows the customer to run operating systems and applications software on cloud infrastructure. The Platform as a Service (PaaS) model delivers a computing platform typically including operating system, programming language execution environment, database, and web server. Network as a Service (NaaS) involves the optimization of resource allocations by considering network and computing resources as a unified whole.

There are numerous sources for researching trends in cloud computing. One such source is SearchCloudComputing.com. The website is an online trade magazine devoted to cloud computing news, analysis and case studies and is a member of TechTarget.com, which is a global network of technology-specific websites. According to SearchCloudComputing, it "provides IT professionals with real-world examples of how cloud computing is being used today" [7]. Staff writer Bill Claybrook stated that in 2013, "Software as a Service (SaaS) will see heavy gains as more customers realize the value of renting software" [7]. In the same article, staff writer Roger Jennings predicts SaaS, PaaS and IaaS combined will grow at a compounded annual growth rate (CAGR) of about 25%" [7]. While these statements do not explain how data center managers will be affected in 2013, they are possibly indicators of market trends that will affect employment.

Long term trends in cloud computing can be found in numerous places as well but perhaps a good source of information is the Bureau of Labor Statistics. The federal government tracks trends in employment rates and makes projections based on those trends. According to the BLS website, "employment of computer and information systems managers is projected to grow 18 percent from 2010 to 2020" [8]. While this projection does not specifically name data center managers, these managers do fall under this heading. One of the factors driving this is cloud computing. According to the BLS, "an increase in cloud computing may shift some IT services to computer systems design and related services firms, concentrating jobs in that industry" [8].

A good example of where data centers could possibly be headed can be found in an article published on the SearchCIO-Midmarket website [9]. According to the executive editor, Christina Torode, "CIOs describe a hybrid data center approach that involves a blend of many strategies: including renting data center space, turning over some applications to a Software as a Service (SaaS) vendor or disaster recovery (DR) to an Infrastructure as a Service (IaaS) provider, and continuing to update and virtualize their own data center infrastructure and develop an internal cloud" [9]. This article is only one of thousands describing future IT solutions involving cloud computing. Data centers still seem to be at the heart of the IT operations even with the propagation of cloud computing.

Two IT staffers for the Oregon's Multnomah County office resigned at the same time, leaving no one to manage their IT service. Since the remaining staff was struggling with the specialized server environment, the County decided to leap into the cloud. Training other IT staffers on server storage, backup administration, recovery and upgrades would have compounded the on-premises software expenses [10]. The County found out that with the infrastructure and application administration offloaded to the cloud, IT could handle most configuration, testing and disaster recovery concerns during a regularly scheduled monthly call.

SNL Financial decided that even though they owned a sizeable internal data center, the company's homegrown existing workflow management application was testing the data center limits. SNL Financial IT department would look at the cloud to see if the department could find one to handle the company's current growth needs and then the company would not have to worry about future equipment update that would be needed [11].

According to the IT research firm, Gartner, Inc., four forces could result in data-center space requirements that will shrink dramatically before the decade is out. These include smarter design, energy efficiency pressures, the realities of high-density environments, and the potential of cloud computing. The focus of the new data center will be on core business services, and, as those services continue to demand more IT resource, the shrinking size of servers and storage will more than offset that growth. With the availability of offloading services to the cloud, management of IT assets will be shifted to the provider.

With the advancement of cloud computing, the responsibilities of a data center manager have changed. Using virtualization on the servers allows for better resource utilization that results in more intelligent computing and better cost management. New monitoring and management tools have made the data center easier to control. The cloud computing and virtualization allows for a more efficient environment with better cooling and management practices. With cloud computing, if growth or a disaster, other outside cloud computing systems could be utilized [12].

According to Cisco, by 2016 total volume of traffic passing through data centers around the world will grow by four times. The annual growth rate of traffic volume of the cloud will increase by 44 percent. The biggest impact of the projected traffic growth will be felt in the data center, where the traffic volume is 8 times greater than in the case of end-users [13]. With increasing server computing capacity and virtualization, multiple workloads per physical server are common in cloud architectures. The results include fewer physical servers to support, more efficient data center operations and more ubiquitous access to networks services and content for consumers and business users.

Consequently, through virtualization and advanced technology enterprise, hosting, and portal datacenters are becoming more efficient [14]. In addition, these efficiencies are correlated with the effective use of the cloud-computing model. Therefore, the size of datacenters is continually expanding due to automation. According to the IDC, the number of U.S. datacenter will decline from 2.94 million in 2012 to 2.89 million by the end of 2016 [15]. However, the IDC also anticipates that the square footage of the existing datacenters will increase from 611.4 million square feet in 2012 to more than 700 million square feet in 2016 [15].

Currently, most datacenters are undergoing a transition from traditional to software-defined datacenter (SDDC). According to VMware, "these types of datacenters are optimized for the cloud era, providing unmatched business agility, the highest SLAs for all applications, dramatically simpler operations, and lower costs", [16]. In addition, this type of datacenter can be adapted to all cloud models. Furthermore, this type of datacenter frees the application layer from the hardware layer allowing for deploying, managing, storing, computing and networking business applications in a cloud computing environment [17]. Overall, the changes in the features and functions of the datacenter will benefit management issues, but require a different skill set than the traditional datacenter in the cloud computing era.

The research shows the growing popularity of all cloud models and services in the datacenter industry. Furthermore, virtualization, automation, and the advancement of technology were noted as the "foundation" for the popularity. The trends discussed, all lead to making datacenters more efficient in size, speed, and capabilities. In addition, the trends suggest a reduction in the amount of datacenters, but conclude that their total size will increase. Overall, the trends confirm that the skill set of a datacenter manager and total responsibilities, due impart to the growing size of the datacenter, will increase.

Moreover, one of the trends mentioned an 18 percent increase in information system managers by 2012 [8]. Therefore, datacenter managers will also be in greater demand. On the other hand, no trend suggests how datacenter managers will adapt in order to serve larger datacenters with more virtualization, automation, size, and other capabilities. Overall, none of the current trends answers the problem that relates to the shortage of datacenter managers due to the growing skill set required and size of datacenters.

III. RUNNING THE NUMBERS

From viewing the numbers previously mentioned, datacenter managers should not worry about losing their jobs but that does not mean these managers will not have to adapt to the ever-changing world of computing. The evidence points toward slightly fewer datacenters but more employees. This seems counterintuitive at first but can be explained as such, if datacenters are becoming cloud centers, that means that datacenter managers are becoming cloud center managers. This helps explains why the BLS projections show an 18% increase in information system managers between 2010 and 2020.

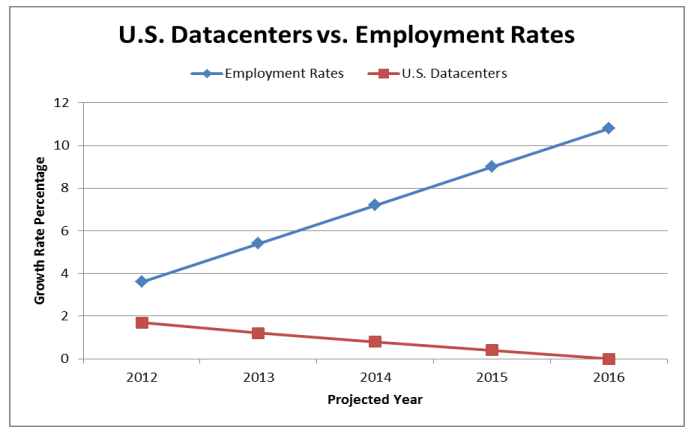

Figure 1. U.S. Datacenters vs. Employment Rates [8] [15]

In Figure 1, a contrast of the total expected growth rate of U.S. datacenters versus the expected employment rate is depicted. In addition, the expected 1.7 percent decrease in U.S. datacenters by 2016 is not affecting the expected employment rate projections from the BLS. Therefore, there is little correlation between the total numbers of datacenters versus the employment rate of datacenter managers.

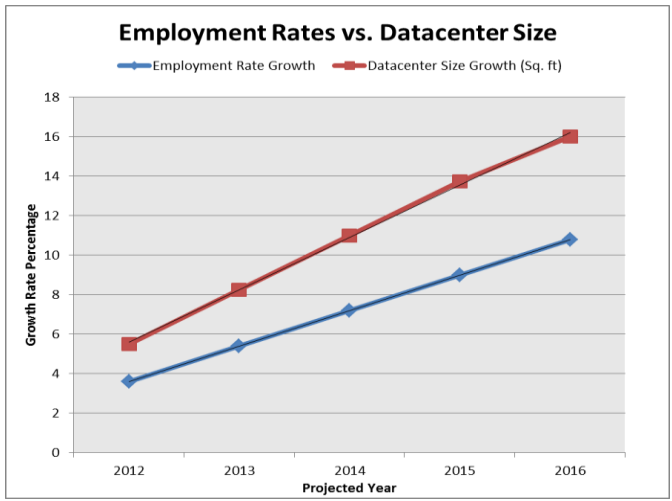

Figure 2. Datacenter Size vs. Employment Rates [8] [15]

In Figure 2, the growth rate of the datacenter is exceeding the employment growth rate. However, this correlation of trends depicts that an increase in datacenter employment is looming. Moreover, the increase in datacenter size is perpetually increasing the need for datacenter managers. Therefore, even though fewer managers are needed per square foot, the overall trend will remain that more datacenter managers will be needed as the total square footage demand of datacenters increases. Overall, the total numbers and square footage of datacenters have different effects on the employment of datacenter managers for various reasons.

The change from datacenter to cloud center means adaptation for datacenter managers not unemployment. Existing managers will need to adapt in a number of areas. Those that are left in legacy datacenters will need to become proficient at incorporating new cloud friendly monitoring and management tools into their existing centers. New datacenter managers will need to integrate these tools from the beginning. Those managers that are on the cloud end of things will have to learn to interface their centers to the needs of an ever-expanding base of customers with varying storage, software and service needs [18] [19].

During the global business recession, the outsourcing of data center work seemed to suffer a setback as organizations took fewer risks and focused on eliminating all but the most necessary functions, including the very managers needed to reengineer IT processes. Now that IT spending is recovering, organizations may begin to shed data centers in favor of cloud-based resources. With improved data center productivity, fewer people are managing more servers, terabytes of storage, and applications. As the cost of operating data centers declines, organizations can afford to invest in more data center capacity. The system support personnel need to know how to build and support the flexible structure that can embrace the cloud while maintaining critical resources in-house [20].

IV. THE BOTTOM LINE

Datacenter managers should not worry about jobs. With the decline in the number of datacenters also comes an increase in the number of cloud-centers. The bottom line is the jobs are still going to be there as the out-pouring of data out paces the efficiencies that are gained in datacenters. As stated earlier in the manuscript, the Bureau of Labor Statistics is speculating that jobs in this field will indeed increase. Cloud facilities will need managers; datacenters will need managers; this explains the projected increase of 18% by the BLS [8]. As long as data continues to grow, which will be the case for the near future; there will be a need for datacenters managers. Those looking for jobs in the management of datacenters should not fear and follow this career path.

In addition, managers need to be concerned with staying current with IT education, so they can manage the latest trends in the profession. Cloud facilities are not taking datacenter manager jobs away, cloud models are requiring the mangers to be more efficient in the way they perform their job. As long as the use of computer technology grows, the need for datacenter managers will continue to grow.

Even though, the projections predict a reduction in the total number of datacenters built until 2016, the employment rate projections are predicted to increase along the same trend line as the total square footage increases. Furthermore, the growth of the datacenter is directly impacted by the use of public and private cloud models to harness a more efficient means of achieving automation. Therefore, the cloud models are spurring the predicted growth in employment rates for datacenter managers.

V. CONCLUSIONS

The demand for data storage is in an ever increasing state. Corporations are finding new avenues to automate the process in order to reduce SLA times, a more efficient means to manage digital content, and reduce the cost of digital storage. The cloud service models, whether it is public, private, etc…, are excellent options due to virtualization, automation, and the advancement of technology. Therefore, the datacenter manager will have to upgrade their skill set to sustain this ongoing development.

On the other hand, the demand for datacenter managers will continue to increase in order to compensate for the increasing demand for data storage. Furthermore, even though an expected decline in the total number of datacenters exists, datacenters are currently increasing in size. This increase in size is due to the use of the cloud models of service. Accordingly, this manuscript compares the expected employment rate projections, with the increase in total projected datacenter size per square foot.

Based on the findings, the expected growth in the employment rate for datacenter managers is spurred by the projected growth in size of the datacenter. However, the findings are limited by the lack of historical data, since newer technologies are fueling the use of the cloud service models. Future research will need to be conducted to verify that the trends stay consistent with the projections. Overall, based on the information available, from the past few years, datacenter manager employment rates in the cloud computing era will continue to increase well into 2016.

WCSIT 3 (3), 65 -69, 2013

ACKNOWLEDGMENT

The authors extend a heartfelt appreciation to Dr. Timur Mirzoev who acted as an expert reviewer for the manuscript.

AUTHORS PROFILE

**Dr. Timur Mirzoev** is an Associate Professor of Information Technology Department at Georgia Southern University, Allen E. Paulson College of Engineering and Information Technology. Dr. Mirzoev heads the Cloud Computing Research Laboratory, Regional VMware IT Academy and EMC Academic Alliance at Georgia Southern University. Some of Dr. Mirzoev's research interests include server and network storage virtualization, cloud systems, storage networks and topologies. Currently, Dr. Mirzoev is holds the following certifications: VMware Certified Instructor, VMware Certified Professional 5, EMC Proven Professional, LefthandNetworks (HP) SAN/iQ, A+.

**Mickey Lewis** has worked in the information technology field in some capacity for the past 11 years. Currently he holds a managerial role at the United Parcel Service in Savannah, GA. He is slated to graduate with honors from Southern Polytechnic State University in the Spring of 2013 with a BSIT.

**David Hillhouse** has worked in the aerospace engineering field for 16 years. He is working towards graduation in July of 2013. His field of study is Information Technology and he is expected to graduate with honors from Georgia Southern University.

**Bruce Benson** has worked in the information technology field for over 30 years. He started as a computer operator, worked as an operations manager, and is currently programming for a medical billing company in Savannah, Ga. He is expected to graduate from Armstrong Atlantic University in December , 2013 with a WBSIT.